\documentclass{jetpl}
\twocolumn

\usepackage{amsmath,amssymb,epsfig,color,pstricks,bm,graphics,alltt}

\usepackage[colorlinks,plainpages=false,linkcolor=black,urlcolor=blue,citecolor=black,pdfpagemode=UseNone,pdfstartview=FitBH]{hyperref}

\definecolor{MyDarkBlue}{rgb}{0,  0.3,  0.9}
\definecolor{MyDarkBlack}{rgb}{0,  0,  0}

\begin{document}

\lat

\title{Electronic and Magnetic Structure of Possible Iron Based 
Superconductor BaFe$_2$Se$_3$}

\rtitle{Electronic and Magnetic Properties of BaFe$_2$Se$_3$}

\sodtitle{Electronic and Magnetic Structure of Presumable Superconductor BaFe$_2$Se$_3$}

\author{$^a$M.\ V.\ Medvedev, $^b$I.\ A.\ Nekrasov,
$^{a,b}$M.\ V.\ Sadovskii\thanks{E-mail: sadovski@iep.uran.ru}}

\rauthor{M.\ V.\ Medvedev, I.\ A.\ Nekrasov, M.\ V.\ Sadovskii}

\sodauthor{Medvedev, Nekrasov, Sadovskii }

\sodauthor{Medvedev, Nekrasov, Sadovskii}

\address{$^a$Institute for Electrophysics, Russian Academy of Sciences, 
Ural Branch, Amundsen str. 106,  Ekaterinburg, 620016, Russia\\
$^b$Institute for Metal Physics, Russian Academy of Sciences, Ural Branch,
S.Kovalevskoi str. 18, Ekaterinburg, 620990, Russia}

\dates{December 2011}{*}

\abstract{
We present results of LDA calculations (band structure, densities of states,
Fermi surfaces) for possible iron based superconductor BaFe$_2$Se$_3$ (Ba123) 
in normal (paramagnetic) phase. Results are briefly compared with similar data
on prototype BaFe$_2$As$_2$ and (K,Cs)Fe$_2$Se$_2$ superconductors.
Without doping this system is antiferromagnetic 
with $T_N^{exp}\sim$250K and rather complicated magnetic structure. 
Neutron diffraction experiments indicated the possibility of two possible
spin structures (antiferromagnetically ordered ``plaquettes'' or ``zigzags''), 
indistinguishable by neutron scattering. Using LSDA calculated exchange 
parameters we estimate Neel temperatures for both spin structures within the 
molecular field approximation and show $\tau_1$ (``plaquettes'') spin 
configuration to be more favorable than $\tau_2$ (``zigzags'').
}

\PACS{71.20.-b, 74.70.-b, 75.10.-b, 75.25.+z}

\maketitle

\section{Introduction}

During the recent years novel iron based high-temperature superconductors 
\cite{kamihara_08} attracted a huge number of both experimental and theoretical 
investigations (for review see~\cite{UFN_90,Hoso_09}).
This flow of papers is still continuing uninterrupted. Major chemical classes
of iron based superconductors are pnictides \cite{kamihara_08} and 
chalcogenides \cite{FeSe} (for comparative study see Ref. \cite{PvsC}).

Recently BaFe$_2$Se$_3$ (Ba123) system \cite{Ba123_1} was proposed as possible 
superconductor (analogous to (K,Cs)Fe$_2$Se$_2$) with experimental indications
of $T_c\sim$11K. However, later study claimed no superconductivity in this
system down to 1.8K \cite{Ba123_2}. Both papers  \cite{Ba123_1,Ba123_2} 
exploited neutron diffraction and discovered Ba123 to be antiferromagnetic 
spin-ladder system with Neel temperature of about 250K, and proposed two 
possible spin configurations. In Ref.~\cite{Ba123_1} these spin configurations 
(antiferromagnetically ordered ``plaquettes'' or ``zigzags'') were reported to 
be equivalent in a sense of diffraction picture and thus indistinguishable.
Ref. \cite{Ba123_2} is rather in favour of ``plaquettes'' pattern.

Most recent investigations of the Ba123 system shown that with doping
(Fe deficiency) it becomes a semiconductor with the 
energy gap about 0.2 eV at room temperature \cite{Lei,Saparov}. 
In Ref.~\cite{Saparov} LSDA calculations were also used to calculate total
energies of phases with different magnetic structures. From these results
it can be seen that more favorable is apparently the ``plaquettes'' 
configuration.

In this work we present LDA calculated electronic structure, densities of 
states and Fermi surfaces for Ba123 system in (high-temperature) paramagnetic 
phase (probably most promising for superconductivity) and compare
the results with those for the previously studied systems, such as 
BaFe$_2$As$_2$ and (K,Cs)Fe$_2$Se$_2$. For antiferromagnetic phase
we use LSDA calculated values of Heisenberg model exchange parameters 
to calculate Neel temperatures for  ``plaquettes'' and ``zigzags'' spin 
structures, to compare relative stability of both structures.
To this end we conclude that  ``plaquettes'' configuration is 
more favorable.

\section{Electronic structure}

We start from crystal structure of Ba123, which belong to
orthorhombic {\em Pnma} space group \cite{Ba123_1}. Main structural motiff of the 
Ba123 system is a two-leg ladder going along $b$-axis. 
This ladder is formed by Fe ions which are surrounded by tetrahedrally
coordinated Se ions. Those ladders form a kind of ``checker board'' structure
perpendicular to $b$-axis. This structure is obviously quite different 
from previously studied BaFe$_2$As$_2$ and AFe$_2$Se$_2$ systems
(see Refs. \cite{PvsC,Nekr2,KFeSe}), which have body centered tetragonal crystal 
structure.

Using this experimentally established crystal structure 
(lattice parameters and all atomic positions used in our calculations
were taken from Ref.~\cite{Ba123_1}) we performed LDA
band structure calculations within the linearized muffin-tin orbitals 
method (LMTO)~\cite{LMTO} using default settings.

In Fig.~1 we present LDA calculated band dispersions (on the right side) and 
densities of states (DOS) (on the left side). Similarly to Ba122~\cite{Nekr2} 
and AFe$_2$Se$_2$ \cite{KFeSe} systems electronic states on the Fermi level and 
around it (from 1 to -2 eV) are mostly formed by Fe-3d orbitals. The Se-4p 
orbitals form bands at energies below -2 eV. Hybridization between  Fe-3d and 
Se-4p states is rather moderate.

\begin{figure}[h]
\includegraphics[clip=true,width=0.5\textwidth]{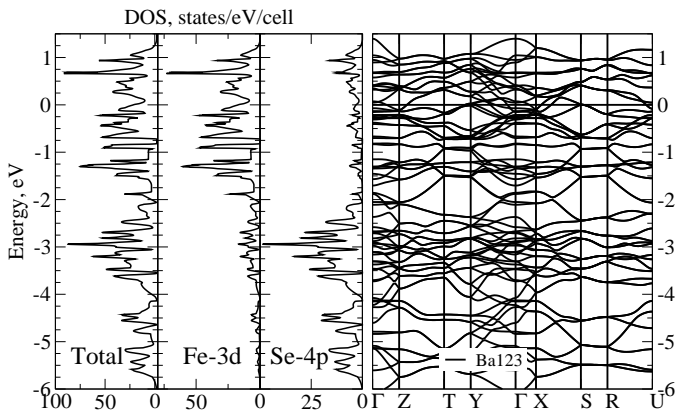}
\caption{Fig. 1. LDA calculated band dispersions and densities of states of 
paramagnetic Ba123. The Fermi level $E_F$ is at zero energy.} 
\end{figure}

Further on in Fig.~2 we show band dispersions of Ba123  in close proximity of
the Fermi level. We can see that these band dispersions are
very much different from typical bands of previously studied pnictides or 
chalcogenides \cite{Nekr2,KFeSe}.
Close to $\Gamma$-point there are two electron pockets and near the Brillouin 
zone border (Y-point), there are three hole-like pockets (see also Fig.~3, 
lower panel).
Close to the Fermi level there are several band crossings and Van-Hove
singularities. Thus Fermi surface topology can be changed rather easy upon 
doping as in AFe$_2$Se$_2$ system \cite{KFeSe}, stressing the potential
importance of doping for superconductivity search.

\begin{figure}[h]
\includegraphics[clip=true,width=0.45\textwidth]{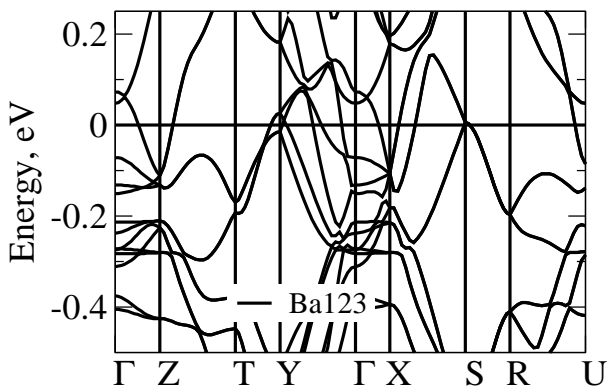}
\caption{Fig. 2. LDA calculated band dispersions in the vicinity of the Fermi 
level for Ba123. The Fermi level is at zero energy.
} 
\end{figure}

In Fig.~3 we present LDA calculated Fermi surfaces (FS) of Ba123.
Overall shape of the Ba123 FS (upper panel of Fig.~4) is very different
from those in typical iron pnictides or chalcogenides \cite{Nekr2,KFeSe}.
First of all it looks pretty three-dimensional and does not have any well
established cylinders. Near $\Gamma$-point we see a jabot-like structure
and near Y-point we observe a kind of ``bottomless pot'' FS sheet.
In the lower panel of Fig.~3 we show FS cross-section for the $k_z=0$ plane.
Fermi sheets closest to $\Gamma$-point are electron like, while those FS sheets 
near Y-point are hole-like.

\begin{figure}[h]
\includegraphics[clip=true,width=0.45\textwidth]{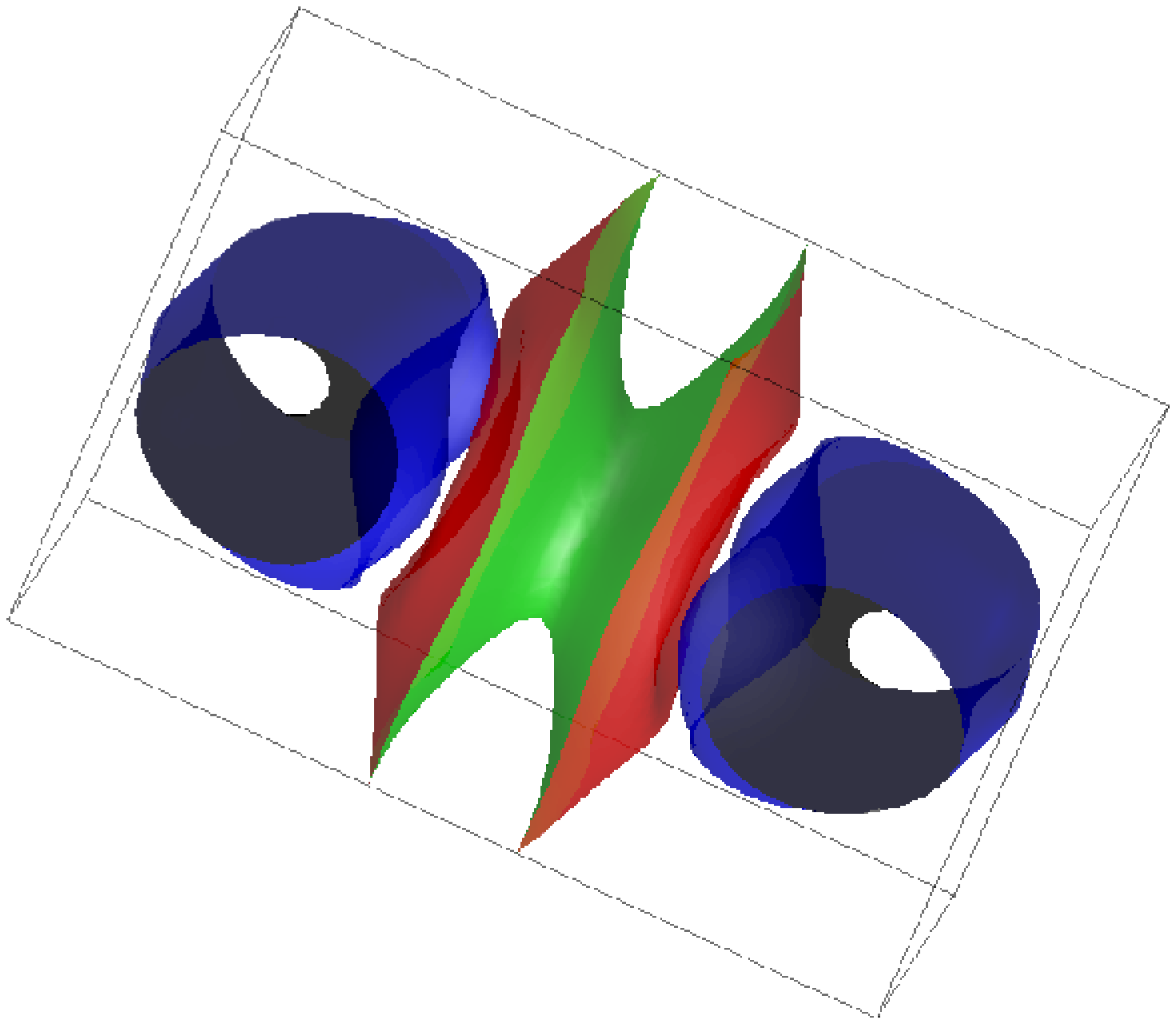}
\includegraphics[clip=true,width=0.45\textwidth]{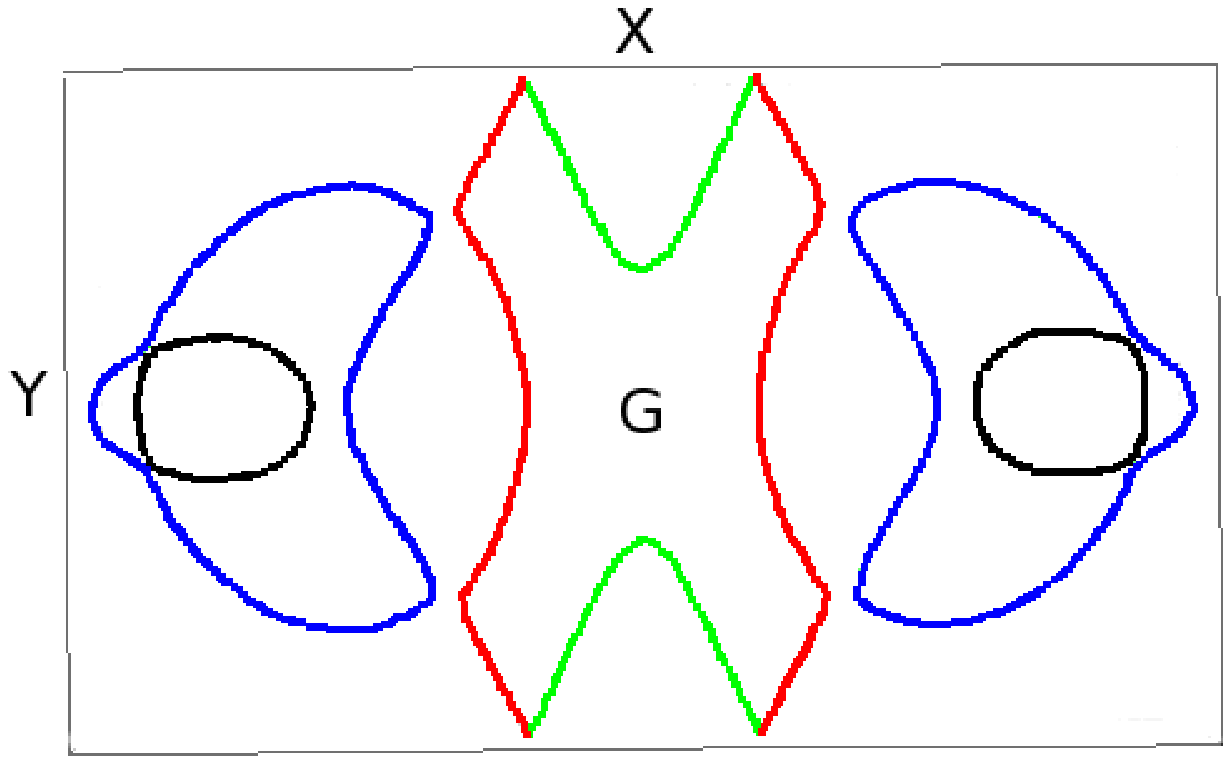}
\caption{Fig. 3. LDA calculated Fermi surface in the  Brillouin zone
(upper panel) and $k_z=0$ plane Fermi surface crossection (lower panel).} 
\end{figure}

\section{Magnetic structure}

\begin{figure}[h]
\centering
$\tau_1$ spin configuration of Ba123 (``plaquettes'')
\vskip 0.25cm
\includegraphics[clip=true,width=0.4\textwidth]{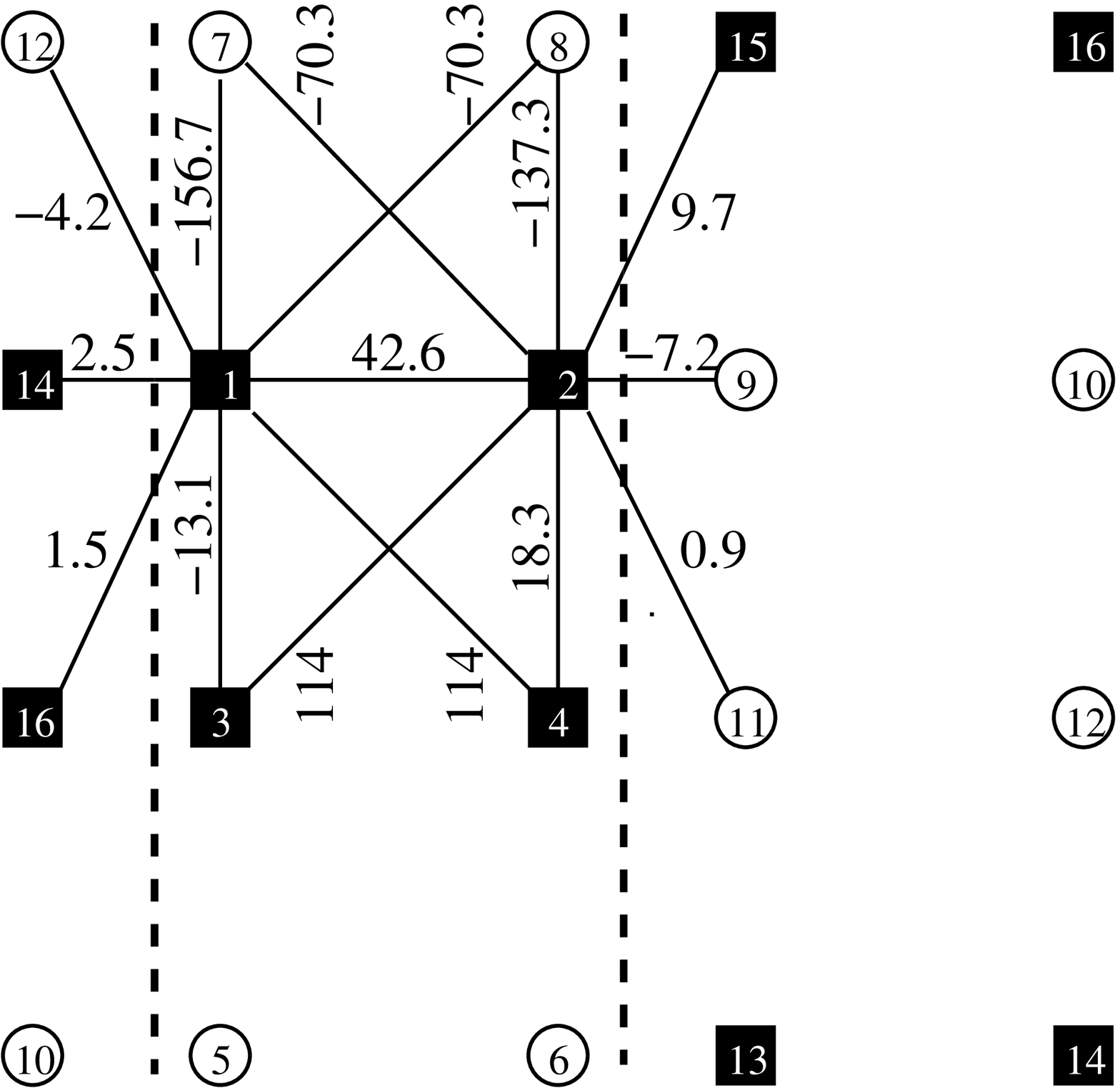}\\
\vskip 0.25cm
$\tau_2$ spin configuration of Ba123 (``zigzags'')
\vskip 0.25cm
\includegraphics[clip=true,width=0.4\textwidth]{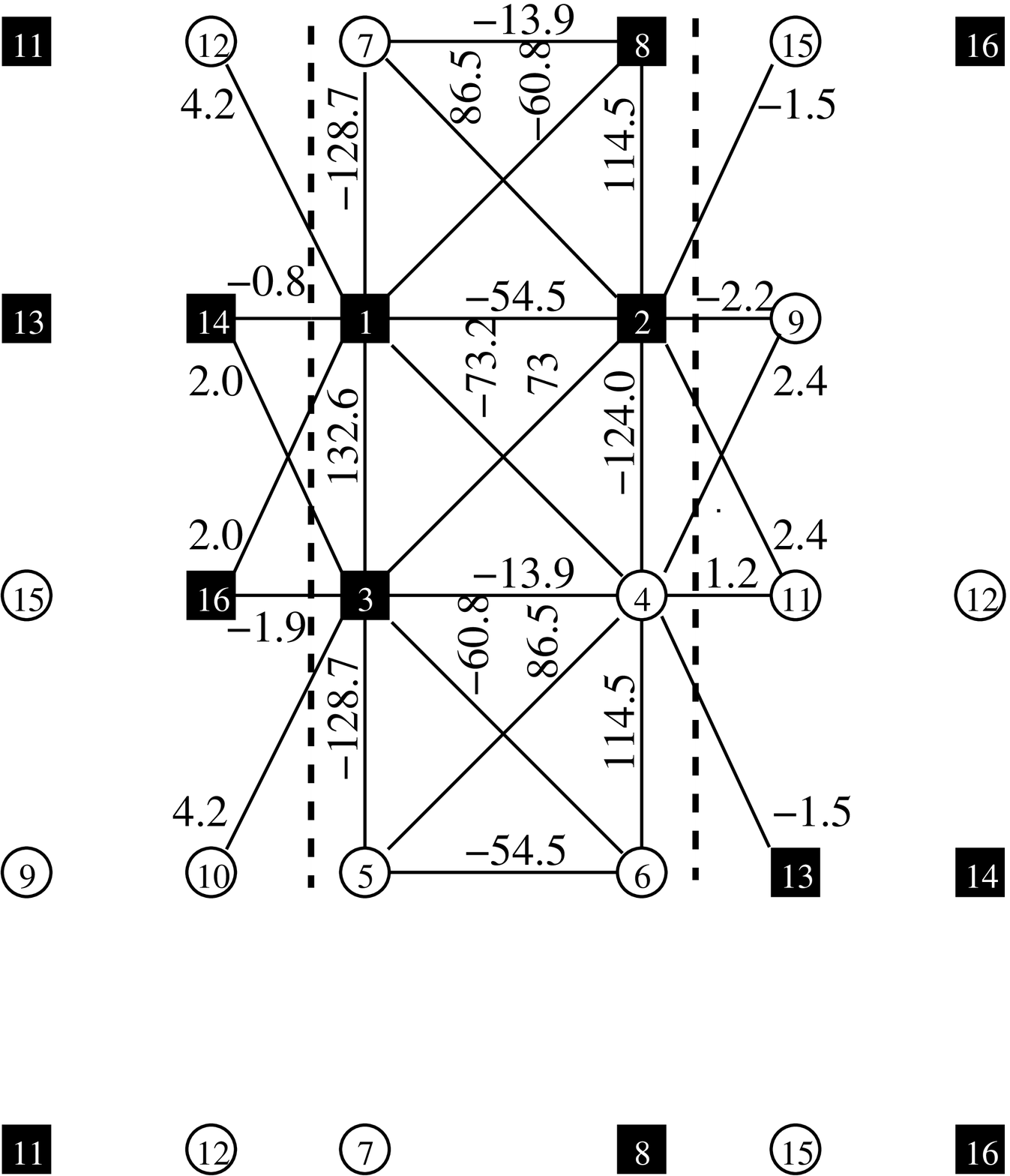}
\caption{Fig. 4. Experimental variants of
AFM spin configurations $\tau_1$ (upper panel) and $\tau_2$ (lower panel) with
LSDA calculated Heisenberg exchange parameter values (K).
``Spin up'' orientation corresponds to circles and ``spin down'' to squares.
Figures in circles and squares enumerate magnetic sublattuces.
Dashed lines separate different ladders.} 
\end{figure}

Neutron scattering experiments \cite{Ba123_1,Ba123_2} on magnetic structure
of Ba123 at $T\ll T_N^{exp}\sim$250K can be interpreted as indicating two
possible types of spin configurations, corresponding to irreducible 
representations $\tau_1$ and $\tau_2$ of the {\em Pnma} space group, 
giving the same diffraction peaks structure.
Since neutron diffraction experiments can not distinguish between these 
magnetic structures, it is tempting to find more stable configuration 
theoretically. One can do that by computation of Neel temperature for both 
configurations within the molecular field approximation, using LSDA calculated 
Heisenberg exchange interaction parameters. 

In case of magnetic structure with several collinear magnetic
sublattices Heisenberg model Hamiltonian for classical spin vectors can be 
written as:
\begin{eqnarray}
H &=&-\frac{1}{2} \sum_{i,n\neq \j,m} I_{i,n,j,m} \vec{S}_{i,n} \vec{S}_{j,m} = \nonumber \\
  &=&-\frac{1}{2} \sum_{i,n\neq \j,m} I_{i,n,j,m} S^2 \vec{e}_{i,n} \vec{e}_{j,m},
\end{eqnarray}
where $I_{i,n,j,m}$ is exchange integral between site $i$ of sublattice $n$
and site $j$ of sublattice $m$, $S$ -- classical spin vector length and
$\vec{e}_{i,n}$ is the unit vector of spin direction. 
Self-consistent equations for thermodynamic
averages of spin $z$-projections $\sigma_{in}\equiv <S^z_{i,n}>$ within
mean-field approach can be linearized near Neel temperature $T_{N}$ 
and written as:
\begin{eqnarray}
T_{N} \sigma_{in} = \frac{S^3}{3} \sum_{jm} I_{in,jm} \sigma_{jm}.
\end{eqnarray}
Due to translation invariance of the crystal the values of $\sigma_{in}=\sigma_n$
are independent of site number in magnetic sublattice. Then Neel
temperature $T_{N}$ is determined by the solution of the full system of
linear equations for $\sigma_n$ for all magnetic sublattices.

The values of exchange integrals for both spin configurations ($\tau_1$ and 
$\tau_2$) in Ba123 antiferromagnet for classical Heisenberg model at $T=0$
were calculated using the method proposed in Ref.~\cite{leip}. The results are
shown by numbers (in K) on bonds in Fig.~4.

Note that both $\tau_1$ and $\tau_2$ structures can be represented as
a set of 16 collinear sublattices with 8 with ``spin up'' and 8 with
``spin down'' orientations. The absolute value of $S$ at $T=0$ on each Fe ion in
all sublattices is the same. However spins on different sublattices are 
connected with their magnetic surroundings by different sets of exchange 
integrals. Thus, at finite temperature the
values $|\sigma_n| = |<S^2_{in}>|$ will be different for different sites.
We restrict ourselves to the account of only nearest and
next nearest exchange bonds of a given ion in the spin-ladder where it belongs
and also coupling of nearest neighbors in adjacent ladders.

For $\tau_1$ spin structure, using the symmetry
of nearest magnetic environment (see Fig. 4, upper panel), we can see that the 
following relations should be fulfilled:
\begin{eqnarray}
 \sigma_{1} = \sigma_{3} = -\sigma_{5} = -\sigma_{7} = -\sigma_{10} = -\sigma_{12} = \sigma_{14}  = \sigma_{16} \nonumber \\
 \sigma_{2} = \sigma_{4} = -\sigma_{6} = -\sigma_{8} = -\sigma_{9}  = -\sigma_{11} = \sigma_{13}  = \sigma_{15}.
\end{eqnarray}
Then the linearized equation for the average value of $z$-projection of 
$\sigma_1$ spin on the magnetic sublattice number 1 becomes:
\begin{eqnarray}
T_{N} \sigma_{1} &=& \frac{S^3}{3} \bigl[ J^0_{1,2}\sigma_2 + J^0_{1,3}\sigma_3 + J^0_{1,7}\sigma_7  + \nonumber \\
+ J_{1,4}\sigma_4 &+& J_{1,8}\sigma_8 + 2\tilde{J}^0_{1,14}\sigma_{14} + 2\tilde{J}_{1,12}\sigma_{12} + 2\tilde{J}_{1,16}\sigma_{16} \bigr],
\end{eqnarray}
where $J^0_{1,2}$, $J^0_{1,3}$ and $J^0_{1,7}$ are nearest exchange parameters 
for selected site of the sublattice 1, with neighbors in 2, 3 and 7 sublattices 
within the ladder, to which the selected site belongs;
$J_{1,4}$ and $J_{1,8}$ are next nearest exchange parameters within the ladder 
to which the selected site belongs;
$\tilde{J}^0_{1,14}$, $\tilde{J}_{1,12}$ and $\tilde{J}_{1,16}$ are 
adjacent ladder nearest exchange parameters.

Using Eq. (3) the Eq. (4) can be reduced to:
 \begin{eqnarray}
T_{N} \sigma_{1} &=& \frac{S^3}{3} \bigl[ \bigl( J^0_{1,3} + J^0_{1,7} + 2\tilde{J}^0_{1,14} - 2\tilde{J}_{1,12}
+ 2\tilde{J}_{1,16} \bigr) \sigma_1 \nonumber \\
&+& \bigl( J^0_{1,2} + J_{1,4} - J_{1,8}\bigr) \sigma_2 \bigr].
\end{eqnarray}
Similar equation for $\sigma_2$ takes the form:
\begin{eqnarray}
T_{N} \sigma_{2} &=& \frac{S^3}{3} \bigl[ \bigl( J^0_{2,1} + J^0_{2,3} - J_{2,7}\bigr) \sigma_1 + \nonumber \\
&+& \bigl( J^0_{2,4} - J_{2,8} - 2\tilde{J}^0_{2,9} + 2\tilde{J}_{2,15} - 2\tilde{J}_{2,11} \bigr) \sigma_2 \bigr].
\end{eqnarray}
Employing the calculated values of exchange parameters given (in K) in Fig.~4,
together with calculated value of magnetic moment on iron $\mu_{Fe}$ = 2.55$\mu_B$
(corresponding classical spin vector value $S$=1.275), one can get from Eqs. (5) 
and (6) the Neel temperature $T_N(\tau_1)$=217K, which is quite close to the 
experimental value of $T_N^{exp}\sim$250K. Note that the account of exchange 
bonds with adjacent ladders $\tilde{J}$ only slightly influences the value of
$T_N$ (without the account of $\tilde{J}$ we get $T_N(\tau_1)$=204K).
However, the presence of $\tilde{J}$ makes symmetry of exchange surrounding
different for sublattices 1 and 2. This significantly influences the values and 
signs of exchange bonds along right and left sides of the spin ladder. 

Antiferromagnetic structure with spin configuration $\tau_2$
is more complicated, however, using magnetic environment symmetry
one can establish the following equalities between average values of 
$z$-projections of spins over all 16 sublattices (see Fig. 4, lower panel).
\begin{eqnarray}
 \sigma_{1} = -\sigma_{5} = -\sigma_{10} = \sigma_{14},~~\sigma_{2} = -\sigma_{6} = -\sigma_{9}  = \sigma_{13}, \nonumber \\
 \sigma_{3} = -\sigma_{7} = -\sigma_{12} = \sigma_{16},~~-\sigma_{4} = \sigma_{8} = \sigma_{11}  = -\sigma_{15}.
\end{eqnarray}
Then following the same line of arguments as for calculation of $T_N(\tau_1)$
(Eqs. (4-6)), one can get four equations for $\sigma_1$, $\sigma_2$,
$\sigma_3$ and $\sigma_4$. Again using the calculated value of
iron local magnetic moment $\mu_{Fe}$ = 2.43$\mu_B$ ($S$=1.215)
and exchange integral values (in K) (see Fig. 4, lower panel),
one can obtain $T_N(\tau_2)$=186K. This value is 36K lower than $T_N(\tau_1)$. 
This tells us that spin configuration $\tau_1$ is, in fact, more favorable.
It is probable that one of the reasons for this result is the presence
of strong antiferromagnetic exchange $J^0_{1,2}=J^0_{5,6}$=-54K
in $\tau_2$ structure. 

\section {Conclusion}

In this work we have studied the band structure, density of states and Fermi 
surfaces of recently discovered possible Fe-based superconductor BaFe$_2$Se$_3$ 
(Ba123). In contrast to typical representatives of iron pnictide and 
chalcogenide systems \cite{Nekr2,KFeSe} and despite rather similar chemical 
composition, Ba123 has completely different crystal structure.
Instead of planes with square lattice of Fe-ions, Ba123 is a spin-ladder system
with two-leg ladders of Fe ions, which form checkerboard pattern in the plane
perpendicular to the ladder direction. Thus, not unexpectedly, the band 
dispersions and Fermi surfaces are quite different from those in
BaFe$_2$As$_2$ and (K,Cs)Fe$_2$Se$_2$. Complicated band structure of 
Ba123 in the vicinity of the Fermi level illustrates the importance of
experiments on different kinds of doping in search for possible 
superconductivity in this system.

For the undoped Ba123 spin-ladder system, there is a problem of correct 
choice among two possible antiferromagnetic spin configurations, not easily
discernible by neutron scattering (``plaquettes'' versus ``zigzags''). We could
distinguish these configurations by calculating the Neel temperature for 
classical Heisenberg model, within the molecular field approximation. 
Corresponding Heisenberg exchange integral values were obtained from
LSDA calculations.  From our estimates we conclude that more favorable
is ``plaquette'' configuration with (calculated) $T_N(\tau_1)$=217K 
(in contrast to ``zigzags'' with calculated $T_N(\tau_2)$=186K). 
This conclusion agrees with experimental evidence of Ref.~\cite{Ba123_2} and
total LSDA energy calculations of Ref.~\cite{Saparov}.

This work is partly supported by RFBR grant 11-02-00147 and was performed
within the framework of the Program of fundamental research of the Russian 
Academy of Sciences (RAS) ``Quantum mesoscopic and disordered structures'' 
(12-$\Pi$-2-1002).

\end{document}